\documentstyle[preprint,epsf,aps]{revtex}

\def\via{{\it via }\thinspace}

\def\hea{He~{\scriptsize I}}
\def\heb{He~{\scriptsize II}}

\begin{document}
\draft

%
\title{\bf Electronic structure of Millerite NiS}
\author{S.~R.~Krishnakumar, N.~Shanthi and D.~D.~Sarma \cite{jnc}}
\address{Solid State and Structural  Chemistry Unit,
Indian Institute of Science, Bangalore, 560012, India}
\date{\today}
\maketitle

\begin{abstract}

We investigate the electronic structure of Nickel sulphide  (NiS)
in the  millerite phase using  electron spectroscopic measurements
and band structure as well as model Hamiltonian calculations.
While band structure calculations are found to be relatively more
successful in describing the experimental valence band spectrum of
this highly conducting phase compared to the hexagonal phase of
NiS, cluster calculations including electron correlation effects
are found to be necessary for the description of certain features
in the experimental spectra, indicating importance of correlation
effects even in a highly metallic system.  The electronic
parameter strengths obtained from these calculations confirm that
the millerite NiS is a highly covalent $pd$-metal. The comparative
study of hexagonal and millerite forms of NiS, provides the
information concerning the evolution of the spectral function in a
$pd$-metal as a function of covalency.

\end{abstract}

\narrowtext

\section{Introduction}

There have been numerous electron spectroscopic investigations of
the  metastable hexagonal phase of nickel monosulphide (NiS),
primarily to understand its  unusual electronic phase transition
\cite{nis_prl,Nakamura1}. However, no electron spectroscopic
studies have been undertaken on the millerite phase of NiS, which
is the {\it stable phase} below 600~K. The resistivity
measurements on the system show that it has a high metallic
conductivity ($\sim 2\times10^{4}$ Ohms$^{-1}$-cm$^{-1}$ at room
temperature). It has been speculated \cite{hulliger@3,benoit@3}
that the millerite phase is diamagnetic with the observed small
magnetic susceptibility  attributed to the presence of
paramagnetic hexagonal modification of NiS. Millerite NiS
crystallizes in the lower symmetry trigonal space group ($R3m$)
with lattice parameters $a=9.589$~\AA\ and $c=3.165$~\AA\
\cite{kolk@3}. The crystal structure of millerite NiS is shown in
Fig.~\ref{mil-stru@3}(a). Here, the Ni atom has five nearest
neighbor S atoms, occupying the corners of a square pyramid. In
this pyramidal geometry (see Fig.~\ref{mil-stru@3}(b)) the Ni
atoms are displaced slightly out of the basal plane, towards the
apical sulphur atom. Though the local coordination of Ni in
millerite is very similar to that in another Ni-S system,
BaNiS$_2$,   millerite has a three-dimensional structure in
contrast to BaNiS$_2$ which is highly two-dimensional in nature.
The Ni-S bond in millerite NiS (shortest $d_{Ni-S}=2.25$~\AA) is
relatively shorter than that in hexagonal NiS
($d_{Ni-S}=2.39$~\AA), NiS$_2$ ($d_{Ni-S}=2.40$~\AA) or BaNiS$_2$
(shortest $d_{Ni-S}=2.32$~\AA) and leads to  stronger
hybridization effects in this system.

There have been very few studies to understand the electronic
structure of this compound. Band structure calculation performed
for the millerite phase of NiS \cite{Raybaud@3} showed the ground
state to be metallic, in agreement with  experimental results.
However, no comparison between experimental and theoretical
results exists and the details of the electronic structure in
terms of the electronic structure parameters are still unknown.

In the present  study, we investigate the electronic structure of
millerite phase of NiS using x-ray   photoemission~(XP) and
Ultra-Violet photoemission~(UP) measurements in conjunction with
{\it ab inito} band structure  as well as parameterized many-body
calculations.  The present  results provide a  consistent and
quantitative description of the electronic structure of the
system. By comparing the hexagonal and millerite phases of NiS, we
study the evolution of the spectral function with changing
correlation effects in the metallic regime.

\section{EXPERIMENTAL}

For the preparation of the millerite samples, polycrystalline
samples of hexagonal NiS was prepared first by solid state
reaction ~\cite{nis_prl,Anzai1}. Hexagonal NiS sample was then
sealed in quartz tube in high vacuum and was maintained at 573~K
for about 2-3 weeks and was cooled slowly to room temperature over
a period of 8~hours, to obtain the millerite phase. The  $x$-ray
powder diffraction measurements confirmed  the phase purity of the
sample.  Spectroscopic measurements were carried out in a combined
VSW spectrometer with a base pressure of 2$\times10^{-10}$ mbar
equipped with a monochromatized Al~K$\alpha$ $x$-ray source and a
Helium discharge lamp.  XP spectroscopic measurements were
performed on the samples with an overall instrumental resolution
of better than 0.8~eV, while UP measurements for He~I and He~II
are performed with an instrumental resolution better than 90 meV
and 120 meV, respectively. The sample surface was cleaned {\it in
situ} periodically during the experiment by scraping with an
alumina file and the surface cleanliness was monitored by
recording the carbon 1$s$ and oxygen 1$s$ XP signal regions.  The
binding energy was calibrated to the instrumental Fermi-level
which was determined by recording the Fermi-edge region from a
clean silver sample.

\section{CALCULATIONS}

Scalar relativistic linearized muffin-tin orbital (LMTO) band
structure calculations have been performed within the atomic
sphere approximation (ASA) for millerite phase of NiS  with the
real crystal structure \cite{kolk@3}. Here, the rhombohedral unit
cell consisting  of 3 formula units was employed. Sphere radii
used for Ni and S were 2.49 and 2.55 a.u., respectively. 17 empty
spheres with sphere radii in the range of 0.9 to 1.65 a.u. were
also used. Convergence was obtained with $s$, $p$ and $d$ orbitals
at Ni and S atomic spheres and $s$ and $p$ for the empty spheres,
with 220~$k$ points in the irreducible part of the Brillouin zone.

Core level and valence band~(VB) spectra were calculated  for
millerite NiS, using a NiS$_5$ cluster as found  in the solid (see
Fig.~\ref{mil-stru@3}(b)), within a parameterized many-body  model
including orbital dependent electron-electron (multiplet)
interactions.   The calculational method has been described in
detail elsewhere \cite{nis_prb,baco_prb}. The calculations were
performed including all  the transition metal 3$d$ and sulphur
3$p$ orbitals. In the  VB calculation, Ni~3$d$ and S~3$p$
contributions to  the  valence band spectra were calculated within
the \textit{same} model and with the \textit{same} parameter
values. As the dimension of the Hamiltonian matrix is large,
Lanczos method was used to evaluate the spectral function and the
calculated one-electron removal spectra were appropriately
broadened to simulate the experimental spectra.  In the Ni~2$p$
core level calculation, Doniach-$\breve{S}$unji$\acute{c}$ line
shape function \cite{DS} was used for broadening the discrete
energy spectrum of the cluster model, in order to represent the
asymmetric line shape of core levels from these highly metallic
compounds; similar asymmetric line shapes are also found in the
other core levels in this system.  In the case of VB calculations,
energy dependent Lorentzian function was used for the lifetime
broadening.  Other broadening effects such as the resolution
broadening and solid state effects were taken into account by
convoluting the spectra with a Gaussian function.  The broadening
parameters were found to be consistent with values used for
similar systems \cite{nis_prb,baco_prb,Dimen@3}.  Since the atomic
cross-sections for the Ni~3$d$ and S~3$p$ states  are vastly
different, it is necessary to calculate a weighted average of
these two contributions to the valence band. The atomic
cross-section ratio \cite{Yeh@3} between S $3p$ and Ni $3d$ states
($\approx$ 0.17) is not appropriate in this context, since
solid-state effects alter this ratio significantly \cite{matrix}.
It was found that S 3$p$/Ni 3$d$ cross-section ratio of
approximately 5.5 times that obtained from the atomic calculations
gives the best result for the valence band calculations.

\section{RESULTS \& DISCUSSIONS}

The total as well as  partial Ni~$d$ and S~$p$ densities of states
(DOS) obtained from the LMTO band structure calculation are shown
in Fig.~\ref{mil-dos@3}. The thick solid line represents the total
DOS, while thin solid line and dashed line show the Ni~3$d$ and
S~3$p$ partial DOS, respectively. Our results are in good
agreement with that of a previously published calculation
\cite{Raybaud@3}. The overall features of the DOS for millerite
phase is similar to that obtained for the paramagnetic hexagonal
phase of NiS \cite{nis_prb}. However, in this case, the
Fermi-level lies in the rising part of the DOS, instead of close
to a minimum in DOS as in the case of hexagonal NiS where an
instability in the Fermi-surface can open up a gap in certain
directions of the Fermi-surface \cite{nis_prl}. This distinction
may partially be responsible for the absence of any phase
transition in the millerite phase. The DOS between -3.5~eV and
1.2~eV is dominated by the Ni~3$d$ contributions. Near the Fermi
energy region, there is a substantial contribution from the S~3$p$
states, due to the strong covalency (Ni~$d$-S~$p$ interactions) in
the system, forming the antibonding states. The DOS in the energy
range -5.5~eV to -3.5~eV has  dominant Ni~3$d$ and S~3$p$
contributions and represents the bonding states of the system. In
the energy range, -8~eV to -5.5~eV,  S~3$p$ contribution is
dominant with a smaller contribution from the Ni~3$d$ states;
these non-bonding states of the S~3$p$ are stabilized in energy
compared to Ni~$d$-S~$p$ bonding states due to strong S-S
interactions, similar to the case of hexagonal NiS and other
sulphides \cite{Raybaud@3}.  The peak at $\sim -1.9$~eV resulting
from the Ni~3$d$ ($t_{2g}$-like states) is shifted to a higher
energy compared to the $t_{2g}$ states of the hexagonal NiS. This
is related to the formation of the direct Ni-Ni bonds in the
millerite phase leading to the stabilization of this phase
compared to the hexagonal phase \cite{Raybaud@3}.

The band structure results are compared with the experimental data
in Fig.~\ref{mil-dos-fit@3}. The partial densities of states of
Ni~3$d$ and S~3$p$ are broadened with the experimental XP
spectroscopic resolution of 0.8~eV and are shown along with
valence band spectra taken with 21.2~eV (\hea), 40.8~eV (\heb) and
1486.6~eV (Al~K$\alpha$) photon energies. Such a comparison
provides an experimental determination of the orbital characters
of the various features in the experimental spectra, since the
photoemission cross sections for Ni 3$d$ and S 3$p$ vary
significantly with the photon energy. The decrease in the
intensity of features B and C in going from 21.2~eV photon energy
to 40.8~eV is explained by the Cooper minimum in the S~3$p$
photoionization cross section at around 50~eV, indicating the
dominance of the S~3$p$ contributions in these two spectral
features, with feature A dominated by the Ni~3$d$ contribution.
From band structure results, the feature B has contributions from
S~3$p$ as well as from Ni~3$d$, while the feature C is dominated
by S~3$p$ states. Although the band structure calculations
reproduce the features B and C rather well, the energy position of
the feature A is not correctly reproduced, with the calculated
peak appearing at about 0.4~eV higher binding energy compared to
the experiment. This discrepancy between the experimental results
and band structure calculations is attributed to the electron
correlation effects within the Ni 3$d$ levels; thus, it appears
that correlation effects continue to have important influence even
for this highly metallic system. This has prompted us to go beyond
the band structure theories and study the electronic structure of
this system using a cluster-model, where the electron correlation
effects are explicitly taken into account.

The Ni~2$p$ core level spectrum for the millerite phase of NiS is
shown in  the inset of Fig.~\ref{mil-ni2p-fit@3} by solid circles.
The spectrum consists of spin-orbit split 2$p_{3/2}$ (852.8~eV
binding energy) and 2$p_{1/2}$ (870.2~eV binding energy) signals
with strong satellite features at about 859.5~eV and 876~eV
binding energies, corresponding to 2$p_{3/2}$ and 2$p_{1/2}$
signals, respectively. The intense satellite features in the
Ni~2$p$ core level spectrum point to the presence of the electron
correlations in the system. The 2$p_{3/2}$ and 2$p_{1/2}$ peaks
show strong asymmetries, similar to hexagonal form of NiS
\cite{nis_prb}.  In order to determine the inelastic scattering
background, we have  performed  electron  energy  loss
spectroscopy  (EELS)  on  these samples, with the same primary
energy  as that of the Ni~$2p$ core level peak. Using a procedure
that have been previously employed \cite{nis_prb,baco_prb},  the
inelastic background  function obtained for millerite NiS is shown
in the inset of Fig.~\ref{mil-ni2p-fit@3} as a solid line. The
background function is a highly structured function with a broad
plasmon loss feature around 876~eV, giving rise to an apparently
stronger satellite intensity for the Ni~2$p_{1/2}$ peak compared
to the satellite intensity accompanying the 2$p_{3/2}$ signal.

We have performed core and valence band calculations within a
single model for a NiS$_5$ cluster. The cluster fragment used for
the calculation is shown in Fig.~\ref{mil-stru@3}(b). Within this
model, we consider only one Ni atom; effects due to the strong
Ni-Ni interactions present in this compound  could not be
considered due to the prohibitively  large basis set involved in
such calculations. For Ni$^{2+}$, the electron-electron
interaction parameters, $F^2_{dd}$, $F^4_{dd}$, $F^2_{pd}$,
$G^1_{pd}$ and $G^3_{pd}$ used were same as that for hexagonal
form of NiS \cite{nis_prb}. In the main panel of
Fig.~\ref{mil-ni2p-fit@3}, we show the calculated core level
spectrum (solid line) including the experimentally determined
inelastic scattering background, superimposed on the experimental
spectrum (open circles) for the parameter set S-Ni $(pd\sigma)$ =
-1.8~eV, $\Delta$ = 1.0~eV and $U_{dd}$ = 4.0~eV. The calculated
spectrum without any broadening effect is presented as a stick
diagram at the bottom of the same figure. The calculated spectrum
matches reasonably well with the experimental spectrum. However,
there are some differences  between the calculated spectrum and
the experimental one; the rising edge of the Ni~2$p_{3/2}$ at
$\sim 852$~eV and the satellite energy region of the 2$p_{1/2}$
peak are not accurately described by the calculation. Such
discrepancies between the experimental and calculated spectrum may
have its origin in the neglect of  the strong metal-metal bonds in
the system or due to the slight differences in the background
function generated from the EELS spectrum and the actual
background in the photoemission spectrum.

The $(pd\sigma)$ values obtained for millerite NiS (-1.8~eV) is
considerably larger than that obtained for other Ni-S systems
like, hexagonal NiS \cite{nis_prb} ($(pd\sigma) = -1.4$~eV),
NiS$_2$ \cite{nis2_prb} ($(pd\sigma) = -1.5$~eV) and BaNiS$_2$
\cite{baco_prb} ($(pd\sigma) = -1.5$~eV). This significant
increase in the $(pd\sigma)$ for the millerite case is related to
the shorter Ni-S bonds in the system compared to the other nickel
sulphides, as mentioned earlier,  leading to a high degree of
covalency. The value of the charge transfer energy ($\Delta =
1$~eV) obtained for the millerite NiS is smaller than the
hexagonal form of NiS \cite{nis_prb} (2.5~eV) and NiS$_2$
\cite{nis2_prb} (2.0~eV). However, a similar value of $\Delta$ has
been observed in another Ni-S system, BaNiS$_2$ \cite{baco_prb}
($\Delta=1.0$~eV); significantly, BaNiS$_2$ has the  same local
geometry around Ni atoms as that in the millerite NiS with NiS$_5$
cluster in a pyramidal arrangement. It has been observed in the
case of divalent nickel oxides \cite{Dimen@3}  that as the
dimensionality of the Ni-O connectivity decreases, the charge
transfer energy decreases, \via a change in the Madelung
potential. Thus, the decrease in the $\Delta$ for millerite phase
compared to hexagonal NiS is also possibly due to a change in the
Madelung potential arising from the decreased local coordination
of Ni atoms. The lower value of $\Delta$ for the millerite phase
compared to the hexagonal phase leads to a higher degree
hybridization mixing of Ni~$d$ and S~$p$ states in the system,
leading to an enhanced covalency. It turns out that the on-site
Coulomb interaction strength, $U$, is very similar between
millerite, hexagonal NiS \cite{nis_prb} and NiS$_2$
\cite{nis2_prb}. This shows that $U$ is not very sensitive to the
local co-ordination or to the metallic/insulating property within
the series of nickel sulphides. According to the ZSA \cite{ZSA}
phase diagram, these parameters (small $\Delta$ and large hopping
strength) place the millerite phase deep inside the $pd$-metallic
regime.

Our calculations suggest that the ground state wavefunction of the
millerite phase consists of 52\%, 42.6\% and 5.4\% of $d^8$,
$d^9\underline{L}$ and $d^{10}\underline{L}^2$ configurations with
a high-spin  $S = 1$ state. The average $d$-occupancy ($n_d$) of
the system is found to be 8.53, which is substantially larger than
that obtained for hexagonal NiS ($n_d = 8.43$). This establishes a
strongly covalent ground state of the millerite phase, compared to
other sulphides studied. In order to understand the origin of
various features in the experimental core level spectrum, we have
analyzed the characters of some typical  final states, marked 1-10
in the main panel of Fig.~\ref{mil-ni2p-fit@3}, in terms of
contributions  from  various configurations ($d^8$,
$d^9\underline{L}^1$ and $d^{10}\underline{L}^2$)  (see
Table~\ref{mil-ni2p-tab@3}). These features can be grouped into
four regions; the main peak region, 852-855~eV (labelled 1-3 in
Fig.~\ref{mil-ni2p-fit@3}), the weak satellite region between
856~eV and 858~eV (labelled 4 and 5), intense satellite region
between 858~eV and 861~eV (labelled 6-8) and the high energy
satellites beyond 863~eV (labelled 9 and 10). The first, third and
fourth group of features have essentially similar characteristics
as those in the hexagonal phase, with dominant
$d^9\underline{L}^1$, significant contributions from all
configurations, and dominant $d^8$ configuration, respectively
\cite{nis_prb}. However, the second group of features, which is
absent in the hexagonal phase of NiS, have almost 90\% of the
contributions coming from the well-screened $d^9\underline{L}^1$
configuration. Almost pure charge transfer nature of these states
is possibly due to the shorter Ni-S distances and a lower value of
$\Delta$.

Within the same model, we have calculated the XP valence band
spectrum of the millerite phase. The calculated spectrum (solid
line) along with Ni~3$d$ (dashed line) and S~3$p$ (dot-dashed
line) contributions for the millerite phase are shown superimposed
on the experimental XP valence band spectrum (open circles) in
Fig.~\ref{mil-vb-fit@3}. The inelastic scattering background is
taken to have an integral energy dependence and is shown as a
dotted line. The Ni~3$d$ contributions to the VB spectrum without
any broadening effects are shown as a stick diagram. The parameter
set used for the VB calculation is very similar to that used for
the core level calculation; however, it was found that a
$(pd\sigma)$ of -1.6~eV instead of -1.8~eV provides a better
agreement with the experimental result. The agreement between the
experimental and calculated spectra is reasonably good over the
entire energy range. The features A and B are reproduced rather
well, however, the feature C as well as the region close to the
Fermi energy (lower binding energy side of the feature A) could
not be described very accurately. This is however, not very
surprising. The feature C is dominated by states arising from S
$p$-S $p$ interactions, as shown by the band structure results
(see Fig.~\ref{mil-dos-fit@3}). Since we do not take this
interaction in to account, and also due to the intrinsic
limitation of a cluster model to account for such band structure
effects, the feature C is completely absent in the results of the
cluster model. It is reasonable to expect the features close to
the $E_F$ to be influenced by substantial Ni-Ni nearest-neighbor
interactions present in this compound, explaining the discrepancy
between the experiment and the results based on the cluster model
neglecting such metal-metal interactions.

The analysis of the ground state wave function shows that the
ground state has 55.2\%, 40.4\% and 4.4\% of $d^8$,
$d^9\underline{L}^1$ and $d^{10}\underline{L}^2$ configurations,
respectively yielding an average occupancy of 8.49, with a
high-spin configuration. The analysis of the final state
configurations for a selected set (marked 1-10 in
Fig.~\ref{mil-vb-fit@3}) of final states have been carried out and
the results are given in table~\ref{mil-vb-tab@3}. The various
features can be grouped into different categories, namely the main
peak region at 1-3~eV binding energy (labelled 1-3, corresponding
to feature A), region B in the range of 4-7~eV (labelled 4-7) and
weaker satellites beyond 7~eV (labelled 8-10 and corresponding to
feature D). The first group of features has the dominant
contribution coming from the well screened $d^8\underline{L}^1$
states with a non-negligible contribution from the poorly screened
$d^7$ states and overscreened $d^9\underline{L}^2$  states. These
contributions are found to be similar to those observed in the
case of hexagonal NiS \cite{nis_prb}. The second group of features
are predominantly contributed by $d^8\underline{L}^1$ and
$d^9\underline{L}^2$ states, while in the case of hexagonal phase
it was dominated by $d^7$ states. The third group of features have
a high degree of $d^7$ character along with contributions from
$d^9\underline{L}^2$ character, suggesting correlation effects
within the Ni $d$ states. These high energy, weak intensity
satellites manifest in the valence band spectrum as a weak tailing
of the spectrum beyond about 7~eV (feature D in
Fig.~\ref{mil-vb-fit@3}).  It is to be noted here that the
characters of these weak intensity features are nearly same as
that of the features appearing between 6 and 8~eV binding energy
in the VB spectrum of hexagonal NiS which have been attributed to
the spectral signature of the lower Hubbard band, persisting in
the $pd$-metallic regime. Hence, the lower Hubbard band features
are shifted to higher binding energies as well as becoming weaker
in intensity in going from hexagonal NiS to millerite NiS,
primarily due to an enhanced hopping strength in the millerite
phase driving the system deeper in to the metallic regime.

Although the presence of the Hubbard bands in the metallic regime
has been predicted by the dynamical mean-field theoretical (DMFT)
calculations, the evolution of these spectral features well inside
the metallic regime has not been studied in detail as a function
of various electronic interaction strengths. A recent DMFT
calculation for the spectral functions of a charge-transfer
insulator near the metal-insulator transition boundary
\cite{watanabe@3} suggests that within the metallic regime, the
spectral signatures of the Hubbard band moves towards the
Fermi-level, collapsing with the coherent states, forming the
metallic ground state of the system with decreasing $\Delta$. On
the other hand, our experimental results  establish that well
inside the $pd$-metallic regime, the Hubbard band is further
stabilized, moving towards higher binding energy region with
increasing metallicity driven by enhanced hopping interaction
strength between Ni~$d$ and S~$p$ states. Further experimental and
theoretical efforts are needed in this direction to have a better
understanding of the evolution of the electronic structure of
these systems.

In conclusion, the millerite phase of NiS has been studied by
means of electron spectroscopic techniques,  band structure and
model Hamiltonian calculations. The band structure calculations
were found to be more successful in describing the experimental
valence band spectrum in comparison to the case of hexagonal NiS,
suggesting a reduced effect of electron correlation in the
millerite phase. This is consistent with the highly conducting
ground state of the millerite phase, in contrast to the
antiferromagnetic and poorly conducting ground state of hexagonal
NiS. However,  calculations including the electron correlation
effects are found to be necessary for the description of certain
features in the experimental spectra, indicating importance of
such interaction effects even for such a highly metallic system.
Thus, it appears that both band structure effects and correlation
effects need to be treated on an equal footing for a complete
description of such systems. The various electronic parameter
strengths obtained from these calculations indicate that the
millerite phase of NiS is a highly covalent metal ($pd$-metal).
From the comparative study between hexagonal and millerite phases
of NiS, the evolution of the spectral functions in a $pd$-metal as
a function of the covalency is discussed.

\section{Acknowledgments}

The authors thank Professor C.~N.~R.~Rao for continued support and
the Department of Science and Technology, and the Board of
Research in Nuclear Sciences, Government of India, for financial
support.  SRK  thanks Dr. P. Mahadevan for helpful discussions.
DDS thanks Dr. M. Methfessel, Dr. A. T. Paxton, and Dr. M. van
Schiljgaarde for making the LMTO-ASA band structure program
available. The authors also thank Professor S.  Ramasesha and the
Supercomputer Education and Research Center, Indian Institute of
Science, for providing the computational facility.

\pagebreak

\begin{figure}

\caption{\label{mil-stru@3} \noindent (a) Schematic representation
of the crystal structure of the millerite  NiS. Dark spheres
represent Ni atoms and grey spheres represent S atoms. (b) The
local structural environment of the Ni atom which is
penta-coordinated with S atoms in the millerite NiS. }
\end{figure}

\begin{figure}

\caption{\label{mil-dos@3} The total density of states (thick
solid line) as well as partial Ni~3$d$ (thin line) and S~3$p$
(dashed line) density of states  obtained from the LMTO band
structure calculations for the millerite phase of NiS. The zero of
the energy scale refers to the Fermi energy. }
\end{figure}

\begin{figure}

\caption{\label{mil-dos-fit@3} Experimental valence band spectra of
millerite NiS  using 21.2 eV (\hea), 40.8 eV (\heb) and 1486.6 eV (XPS)
photon energies along with the LMTO band structure partial DOS for Ni~3$d$
and S~3$p$, broadened by the experimental XP resolution function. Various
features in the spectra are shown by the vertical lines and are labeled as
A, B and C (see text).}
\end{figure}

\begin{figure}

\caption{\label{mil-ni2p-fit@3} The Ni~2$p$ core level spectrum of
the millerite  NiS (solid circles) along with the generated
inelastic scattering background function from the EELS spectrum at
the same primary beam energy are  shown in the inset. In the main
panel, experimental Ni~$2p$ spectrum (open circles) along with the
calculated spectrum (solid line) for  millerite NiS obtained from
the cluster calculation are shown.  Various final states of the
cluster calculation and the corresponding intensity contributions
without any broadening are shown as the stick diagrams. }
\end{figure}

\begin{figure}

\caption{\label{mil-vb-fit@3} The experimental VB spectrum (open
circles) along with the calculated spectrum (solid line), Ni~$3d$
component (dashed line), S~$3p$ component (dot-dashed line) and
the integral background (dotted line) are shown for  millerite
NiS. The final states of the calculation and the corresponding
intensities without any broadening are shown as the energy stick
diagrams. }
\end{figure}

\begin{table}
\caption{\label{mil-ni2p-tab@3} Contributions from various
configurations in the final states of the Ni~2$p$ core level
photoemission in millerite NiS.  The peak numberings correspond to
the labels indicated in Fig.~\ref{mil-ni2p-fit@3}; the
corresponding binding energies (BE) in eV are also shown.}
\vspace{2ex} \small
\begin{tabular}{||c||c|c|c|c|c|c|c|c|c|c||}
Peak no. & 1 & 2 & 3 & 4 & 5 & 6 & 7 & 8 & 9 & 10\\
BE & 852.9 &  853.7  & 854.3  & 857.2 & 857.4 & 859.0 & 859.7 & 860.7 & 864.8 & 866.4 \\ \hline\hline

$d^8$ & 16.48 & 12.57 & 10.06 & 4.77 & 6.35 & 26.67 & 28.96 & 22.91 & 60.43 & 72.18 \\ \hline

$d^9\underline{L}^1$ & 57.35 & 58.53 & 56.63 &  90.58 & 87.05 & 44.58  & 29.89 & 33.25 &
30.84 & 23.81 \\ \hline

$d^{10}\underline{L}^2$  & 26.17 &  28.90 & 33.41 &  4.65 & 6.60 & 28.75 & 41.15 & 43.84 & 8.73 & 4.01 \\
\end{tabular}
\end{table}

\begin{table}

\caption{\label{mil-vb-tab@3}Contributions from various
configurations in the final states of valence band photoemission
in millerite NiS.  The peak numberings correspond to the labels
indicated in Fig.~\ref{mil-vb-fit@3}; the corresponding binding
energies (BE) in eV are also shown.} \vspace{2ex} \small
\begin{tabular}{||c||c|c|c|c|c|c|c|c|c|c||}
Peak no. & 1 & 2 & 3 & 4 & 5 & 6 & 7 & 8 & 9 &  10 \\

BE & 1.4 & 1.67&  2.18 & 4.40 & 4.83  & 5.12 & 6.23 &  7.64 & 8.33 & 8.57 \\ \hline\hline

$d^7$ & 16.70 & 14.64 & 9.34 & 0.94  & 1.69 & 0.90 & 4.3 & 39.66 & 32.55 & 26.98 \\ \hline

$d^8\underline{L}^1$ & 53.93 & 53.70 & 52.81 & 66.49 & 64.14 & 71.58 & 59.25 & 10.12 & 13.39 & 17.56 \\ \hline

$d^9\underline{L}^2$ & 26.90 & 28.96 & 34.36 & 31.99 & 33.37 & 26.98 & 35.68 & 41.44 & 38.94 & 41.08 \\ \hline

$d^{10}\underline{L}^3 $ & 2.47 & 2.70 & 3.49  & 0.58 & 0.80 & 0.54 & 0.77 & 8.78 & 15.12 & 14.38 \\
\end{tabular}
\end{table}

\end{document}